\documentclass[aps, pra, reprint, floatfix, showpacs, footinbib]{revtex4-1}
 \usepackage{graphicx}
 \usepackage{amsmath,amssymb}
\begin{document}
\title{Statistical product distributions for ultracold reactions in external
fields}

\author{Maykel L.~Gonz\'alez-Mart\'{\i}nez}
\email{maykel.gonzalez-martinez@u-psud.fr}
\affiliation{Laboratoire Aim\'e Cotton, CNRS,
             Universit\'e Paris-Sud, ENS Cachan,\\
             B\^{a}t.\ 505, Campus d'Orsay,
             91405 Orsay, France}
\author{Olivier Dulieu}
\affiliation{Laboratoire Aim\'e Cotton, CNRS,
             Universit\'e Paris-Sud, ENS Cachan,\\
             B\^{a}t.\ 505, Campus d'Orsay,
             91405 Orsay, France}
\author{Pascal Larr\'egaray}
\affiliation{Institut des Sciences Mol\'eculaires, CNRS,
             Universit\'e Bordeaux 1,
             UMR 5255,
             33405 Talence, France}
\author{Laurent Bonnet}
\affiliation{Institut des Sciences Mol\'eculaires, CNRS,
             Universit\'e Bordeaux 1,
             UMR 5255,
             33405 Talence, France}

\date{\today}
\begin{abstract}
\noindent
The main limitation of most ultracold chemistry studies to date is the lack of
an analysis of reaction products.  Here, we discuss the first generally
tractable, rigorous theoretical framework for computing statistical
product-state distributions for ultracold reactions in external fields.  We show
that fields have two main effects on the products of a statistical reaction,
by: (1) modifying the product energy levels thus potentially reshaping the
product distributions; and/or (2) adding or removing product states by changing
the reaction exothermicity.  By analyzing these effects and the strength of the
formalism to distinguish between different reaction mechanisms in benchmark
reactions involving $^{40}$K and $^{87}$Rb species, we argue that statistical
predictions will help understanding product formation and control, and lead
developments to realize the full potential of ultracold chemistry.
\end{abstract}


\maketitle
\section{Introduction}
\label{sec:introduction}
Recent advances in producing and trapping species at temperatures below 1~mK
have made ultracold chemistry a reality.  For the first time, reactions can be
explored in the fully quantum regime where resonances, tunneling, quantum
degeneracy and other non-classical effects dominate the reaction dynamics.
In addition, external fields may be used to favor or disfavor typical and exotic
reaction mechanisms thus allowing their scrutiny with unprecedented detail, to
``tailor'' interactions and study the consequences of confinement upon
reactivity.  Fields can also be used to trap the ultracold species, drastically
enhancing interrogation times and making it possible to probe the effects of the
weakest interactions.  Since 2010, pioneering experiments at JILA, Colorado,
started addressing many of these issues in reactions involving mixed ultracold
samples of K, Rb and KRb \cite{SOspelkaus:10b,K-KNi:10,MHGdeMiranda:11a} by
inferring reaction rates through the measurement of reactants losses.

Despite all key findings at JILA, and being commonly acknowledged as one of the
most important applications for ultracold species
\cite{RVKrems:08,DJNesbitt:12}, low-temperature chemical reactions are most
often considered for their role as \emph{obstacles} to the stability of quantum
gases \cite{PSZuchowski:10b,JNByrd:10,MTomza:13a,BZhu:14a}.  This seems natural,
since the main current experimental focus is on the production of ultracold
molecules which reactions obstruct.  Nevertheless, the problem of reaction rates
and their dependence on external fields have attracted much interest;
Qu\'em\'ener and Bohn
\cite{K-KNi:10,GQuemener:10b,GQuemener:10c,MHGdeMiranda:11a}, Idziaszek,
Julienne \emph{et al.}\ \cite{ZIdziaszek:10a,ZIdziaszek:10b,KJachymski:13a},
and Gao~\cite{BGao:10a,BGao:11a} developed quantum models to interpret the rates
measured at JILA, and discussed ``universal'' features in ultracold reactions.

Yet, the study of a chemical reaction goes necessarily beyond a description of
its rate into that of its products: The question of product-state distributions
and state-to-state cross sections is among the biggest experimental and
theoretical challenges in the field, with the potential to bring ultracold
physics into the realm of chemistry \cite{DJNesbitt:12}.  Product distributions
are far more sensitive than reaction rates to finer details of the dynamics, and
would provide an integral view and deeper understanding of the underlying
physics.  In addition, even if neutral products are quite difficult to detect,
ionic products may be more easily monitored in experiments on ultracold
ion-neutral \cite{FHJHall:11} or Penning-ionization reactions
\cite{JJankunas:14a}, and aid understanding radiative molecular formation,
charge transfer and several other mechanisms.  The issue seems specially timely
for the first product distributions were very recently measured for the closely
related process of 3-body recombination in an ultracold atomic gas
\cite{AHarter:13}.

A theory that accounts for external fields is essential as these are ubiquitous
in ultracold experiments: fields are used for trapping the (ultra)cold species
and/or controlling their interactions.  In principle, a full description of
low-temperature reactions can be derived from the quantum formalism for chemical
reactions in electromagnetic fields developed by Tscherbul and Krems
\cite{TVTscherbul:08c}.  However, such rigorous quantum-mechanical calculations
are not currently tractable for most cases of experimental interest.  This is
partly due to the huge number of ro-vibrational states involved in
low-temperature collisions of relatively heavy species
\cite{MMayle:12,MMayle:13a}.  There is, however, a more fundamental reason for
this: Ultracold reactions span the most widely-different energy, length and time
scales explored in chemistry to date, and hence pose an unprecedented challenge
to chemical collision theory.

Nonetheless, Mayle and coworkers \cite{MMayle:12,MMayle:13a} have shown that
total reaction rates, and answers to a number of important questions regarding
the statistics of global properties of such processes, may still be inferred
using arguments from Rice-Ramsperger-Kassel-Marcus (RRKM) theory, the random
matrix theory of nuclear scattering, and multichannel quantum defect theory
(MQDT).  The first empirical indication of such statistical behavior was
recently reported for ultracold collisions of Er atoms \cite{AFrisch:14}.

In this paper, we set up the first theoretical framework for calculating
statistical product-state distributions and state-to-state cross sections for
ultracold reactions in external fields, by connecting traditional statistical
theories \cite{WHauser:52,WHMiller:70b} and ultracold collision theory
\cite{RVKrems:book09}.  We show that such product distributions may be used as
benchmarks for the founding assumptions of the theory, and hence provide
fundamental tests for the statistical arguments of Mayle \textit{et
al.}~\cite{MMayle:12,MMayle:13a}.  Moreover, they allow us to critically
evaluate possible departures from statistical behavior, both qualitatively and
quantitatively, being thus a powerful tool to rationalize any kind of reaction.

\section{Theory}
\label{sec:theory}
\subsection{Statistical formalism in external fields}
\label{sec2:statistical_formalism_in_fields}
We consider a reactive collision between two species, M$^\alpha_1$ and
M$^\alpha_2$, which yields two products, M$^\beta_1$ and M$^\beta_2$, in the
presence of an external field.  Here, $\alpha$ and $\beta$ loosely refer to
reactants and products; in what follows, $\alpha$ is also used to represent the
set of quantum numbers needed to specify the internal states of the reactants,
while $\beta$ is its analogous for the products.

Product-state distributions are obtained from the reaction cross section
$\sigma_\mathrm{r}$, with the probability density associated to an observable
$X$, $\mathcal{P}(X) =
\frac{1}{\sigma_\mathrm{r}}\frac{\partial\sigma_\mathrm{r}}{\partial X}$
\cite{RDLevine:book05}.  In general, the only conserved quantity in an external
field is the projection of the total angular momentum on the field axis, $M$,
and the total reaction cross section can be obtained by adding all contributions
from state-to-state cross sections, $\sigma_\mathrm{r} = \sum_{M \alpha \beta}
\sigma^M_{\alpha\rightarrow\beta}$.  The rigorous quantum-mechanical expression
for the state-to-state cross section from reactant state $\alpha$ to product
state $\beta$, at a given $M$, energy $E$ and field strength $F$ is
\cite{TVTscherbul:08c}
\begin{eqnarray}
 \sigma^M_{\alpha\rightarrow\beta}(E,F) &=&
  \frac{\pi \hbar^2 g_\alpha}{2 \mu_\alpha (E-E_\alpha)} \nonumber \\
  &&\hspace{-10mm}
   \times \sum_{L_\alpha M_{L_\alpha}} \sum_{L_\beta M_{L_\beta}}
   \left| S^M_{\alpha L_\alpha M_{L_\alpha};\beta L_\beta M_{L_\beta}}(E,F)
   \right|^2,
 \label{eq:sigma_atob}
\end{eqnarray}
where $g_\alpha$ is a degeneracy factor that equals 2 if the reactants are
indistinguishable and 1 otherwise, $\mu_\alpha$ is the reactants' reduced mass,
$E_\alpha$ ($E_\beta$) is the species energy in state $\alpha$ ($\beta$), and
$L_\alpha$ ($L_\beta$) is the space-fixed orbital angular momentum of the
reactants (products) with projection $M_{L_\alpha}$ ($M_{L_\beta}$) on the field
axis.  The sums over the absolute squares of $S$-matrix elements define the
transition probability $P^M_{\alpha \rightarrow \beta}$ from state $\alpha$ to
$\beta$.  Our model becomes \emph{statistical} for we assume that reaction
always proceeds through complex formation, whose dynamics renders the reactant
and product channels statistically independent.  Following Hauser and Feshbach
\cite{WHauser:52}, or Miller \cite{WHMiller:70b}
\begin{equation}
  P^M_{\alpha \rightarrow \beta}
   \approx p^M_\alpha\; p^M_\beta/\sum_\gamma p^M_\gamma,
 \label{eq:PMab_stat}
\end{equation}
where the explicit dependence on $E$ and $F$ has been omitted, as in what
follows.  The $p^M$ quantities are capture probabilities---\textit{i.e.},
the probability of complex formation, for a given $M$, when the species collide
in a specific state.  The sum in the denominator runs over all energetically
accessible reactant and product channels, hence the ratio
$p^M_\beta/\sum_\gamma p^M_\gamma$ is the fraction of collision complexes that
dissociate into product state $\beta$.  Eq.~\eqref{eq:PMab_stat} is statistical
for all capture probabilities are considered uncorrelated and is seen to satisfy
the principle of detailed balance.  In a further approximation, our reasoning
may be extended to include inelastic processes, thus treating all
\emph{quenching} events statistically---although nonreactive scattering is less
likely to involve complex formation.

A distinctive feature of ultracold experiments is the possibility to fully
control the reactants' initial states, thus we fix $\alpha$ in what follows.
The statistical state-to-state reaction cross section may be written as
\begin{equation}
 \sigma_{\alpha \rightarrow \beta, \mathrm{r}}
        = \frac{\pi \hbar^2 g_\alpha}{2 \mu_\alpha (E-E_\alpha)}
          \sum_M p^M_\alpha \frac{p^M_\beta}{\sum_\gamma p^M_\gamma},
 \label{eq:sigma_ab,r}
\end{equation}
from which the total reaction cross section $\sigma_{\alpha, \mathrm{r}}$ is
found by summing over all possible product states $\beta$, and the related total
reaction rate $k_{\alpha, \mathrm{r}} \equiv
[2(E-E_\alpha)/\mu_\alpha]^{1/2} \sigma_{\alpha, \mathrm{r}}$.  Quantities for a
specific temperature are obtained by averaging over the corresponding Boltzmann
distribution.  If reaction occurs, the probability density corresponding to an
observable $X$ in the products reads
\begin{equation}
 \mathcal{P}_\alpha(X) =
  \left( \sum_M p^M_\alpha
   \frac{\sum_\beta p^M_\beta}{\sum_\gamma p^M_\gamma} \right)^{-1}
  \left( \sum_M p^M_\alpha
   \frac{\sum_\beta \frac{\partial p^M_\beta}{\partial X}}
        {\sum_\gamma p^M_\gamma} \right),
 \label{eq:PaX}
\end{equation}
where the first term in parenthesis acts like a proportionality constant.  The
form of this distribution is readily visualized by recognizing that the sum over
partial derivatives $\sum_\beta (\partial p^M_\beta/\partial X) = \delta(X)
\sum^{\{X\}}_\beta p^M_\beta$, where $\delta(X)$ is a Dirac $\delta$-function
and the sum includes only product states compatible with the value $X$.
Contributions from different $M$ values combine with different relative weights
that approach the capture probabilities in the entrance channels if many product
states are available.

Equations~\eqref{eq:sigma_ab,r} and \eqref{eq:PaX} provide detailed statistical
predictions for the observables of a chemical reaction in an external field.
In the ultracold regime, kinetic energies in the entrance/elastic (and possibly
inelastic) channels are extremely low, hence a quantum-mechanical description is
crucial for calculating the corresponding capture probabilities.  By comparison,
the kinetic energies available to the products are much larger and a simpler
(semi)classical description should be appropriate.  The issue of capture in the
ultracold entrance channels has received much attention in the last few years,
mainly to describe the loss rates measured at JILA; various models have been
proposed \cite{SOspelkaus:10b,K-KNi:10,ZIdziaszek:10a,BGao:10a,GQuemener:10b,
MHGdeMiranda:11a,AABuchachenko:12a,AABuchachenko:12b,KJachymski:13a} that are
valid under specific conditions.  In addition, we have extended time-independent
\cite{EJRackham:01,TGonzalez-Lezana:07} and time-dependent methods
\cite{HGuo:12}, and a variety of semiclassical and classical models, to make it
possible to evaluate capture probabilities in external fields also for reactions
at higher temperatures \cite{MLGonzalez-Martinez:14c}.  The statistical
formalism is quite flexible as capture models may be chosen among all of the
above, or new models developed, depending on the conditions of the experiment
and desired level of sophistication.

\subsection{Implementation details}
\label{sec2:implementation_details}
The calculations for Eqs.~\eqref{eq:sigma_ab,r} and \eqref{eq:PaX} are performed
in two main steps, which are described below.

First, quantum states are defined for all reactants and products by
diagonalizing the Hamiltonians of the corresponding isolated species.  For each
species M$^\gamma_k$, every such state is associated an energy $E_{\gamma,k}$
and projection of the total angular momentum $m_{\gamma,k}$.  Quantum
numbers/labels are additionally associated to each state from the
eigenfunction(s) with the largest contribution in the chosen basis---which are
later used to produce quantum-state/label distributions.  In addition, effective
electric dipole moments are obtained for each state from the expectation value
of the electric dipole operator.

Secondly, for each projection of the total angular momentum $M$, a capture
probability $p^M_\gamma$ is calculated for every combination of reactant/product
states that satisfies:~(1) it corresponds to an asymptotically open channel,
$E \ge E_{\gamma,1} + E_{\gamma,2}$; and (2) the projection of the total angular
momentum is conserved, $M = m_{\alpha,1} + m_{\alpha,2} + M_{L_\alpha} =
m_{\beta,1} + m_{\beta,2} + M_{L_\beta}$.  The latter involves completing the
definition of channels by adding an orbital angular momentum $L_\gamma$, and
associated projections $M_{L_\gamma}  = -L_\gamma, \ldots, L_\gamma$, to each
combination of reactants/products states.  In the case where only magnetic
fields are present, an additional constraint is imposed based on the
conservation of the total parity---calculated for each channel by multiplying
the parities of the individual species (computed in the first step) and
$(-1)^{L_\gamma}$.  The capture models are chosen on the basis of the physical
context and open channels are added until convergence is achieved.  This whole
step is repeated until convergence is achieved with respect to $M$.

\section{Results and discussion}
\label{sec:results+discussion}
To illustrate our formalism, we focus on benchmark ultracold exoergic reactions
involving $^{40}$K, $^{87}$Rb and their diatomic combinations.  Reactive
collisions between these species are barrierless, with exothermicities $\Delta
E$ ranging from about 10 to 200~cm$^{-1}$, and proceeding through wells up to
8,000~cm$^{-1}$ deep \cite{ERMeyer:10,JNByrd:10}.  Such large energy differences
are usually an indication of strong interactions among internal degrees of
freedom (DOFs) of the intermediate complex.  Strong interactions, a large number
of states associated to the intermediate complex and a relatively few open exit
channels most often lead to long-lived intermediates and statistical behavior.
These are also the conditions for ergodic dynamics in classical phase space,
recently studied by Croft and Bohn in ultracold molecular collisions
\cite{JFECroft:14a}.

Quantum states are computed with the Hamiltonians and parameters in
Refs.~\cite{NIST:atomic,JAldegunde:08,JAldegunde:09}.  Capture probabilities in
the entrance channels are evaluated with a Wentzel-Kramers-Brillouin (WKB)
tunneling model for reactions in magnetic fields, and the adiabatic variant of
the quantum threshold model (QTM) \cite{GQuemener:10b} for those in electric
fields.  A semiclassical model based on phase-space theory (PST)
\cite{JCLight:64,PPechukas:65,EENikitin:65,CKlotz:71}, suitable to account for
long-range dispersion interactions, \textit{cf.},
Refs.~\cite{PLarregaray:06,PBargueno:07a,PBargueno:08a,FDayou:08,MJorfi:09}, is
used for capture probabilities from the products.  In general, capture in
barrierless reactions is dominated by long-range interactions, which are
accounted for by the chosen models.  Nevertheless, if necessary, short-range
effects can be included in the statistical formalism by calculating capture
probabilities using, \textit{e.g.}, a quantum-mechanical
\cite{EJRackham:01,HGuo:12} or quasi-classical approach \cite{FJAoiz:07a}; such
studies require a detailed knowledge of the interaction potentials, which
usually involve computationally expensive \textit{ab initio} calculations.
Further details are described in the Appendix.

\subsection{Global properties}
\label{sec2:global_properties}
We first consider \emph{global} properties, and remark an important consequence
of summing over $\beta$ in Eq.~\eqref{eq:sigma_ab,r}: In the common case where
many more product than reactant states are available, $\sum_\beta
p^M_\beta/\sum_\gamma p^M_\gamma \rightarrow 1$, and global quantities depend
exclusively on the capture probabilities for the \emph{entrance} channels.  Such
observables thus provide a ``one-sided'' description of the process and its
governing dynamics, and contain no information on how (or even which) products
are formed in the reaction.

In cases with many more product than reactant states, statistical predictions
will be similar to those based on the ``universal'' assumption
\cite{ZIdziaszek:10a}---in which the wavefunction is considered to be fully
absorbed once the short-range, chemical region is reached.  However, statistical
and universal predictions are not necessarily indistinguishable.  For instance,
ultracold isotopic-substitution reactions would have very small exothermicities
and only a few available product states, thus global statistical and universal
predictions could differ.  Moreover, the requirement for universality implies
nothing about the intermediate dynamics thus universal product distributions
(not derivable from the ``universal'' methodology at present) would not
necessarily agree with statistical predictions: The statistical assumption
imposes a more stringent constraint, namely that capture probabilities in the
entrance and product channels are statistically independent.

\begin{figure}[!t]
 \includegraphics[width=86mm]{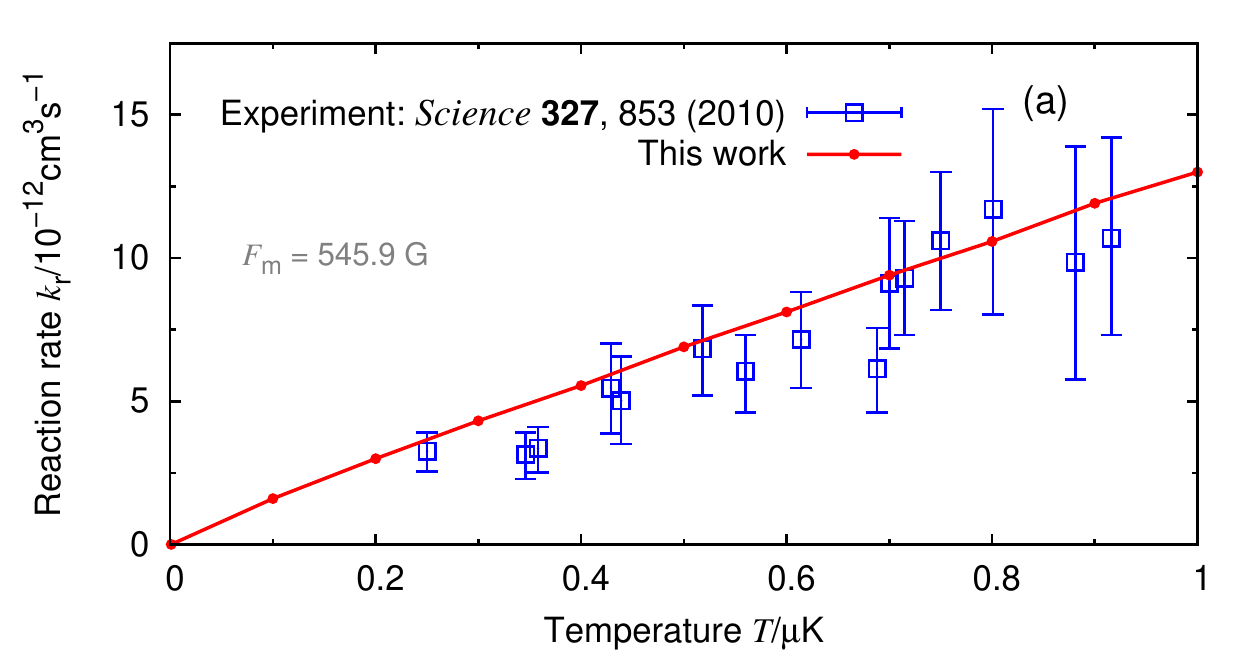}
 \includegraphics[width=86mm]{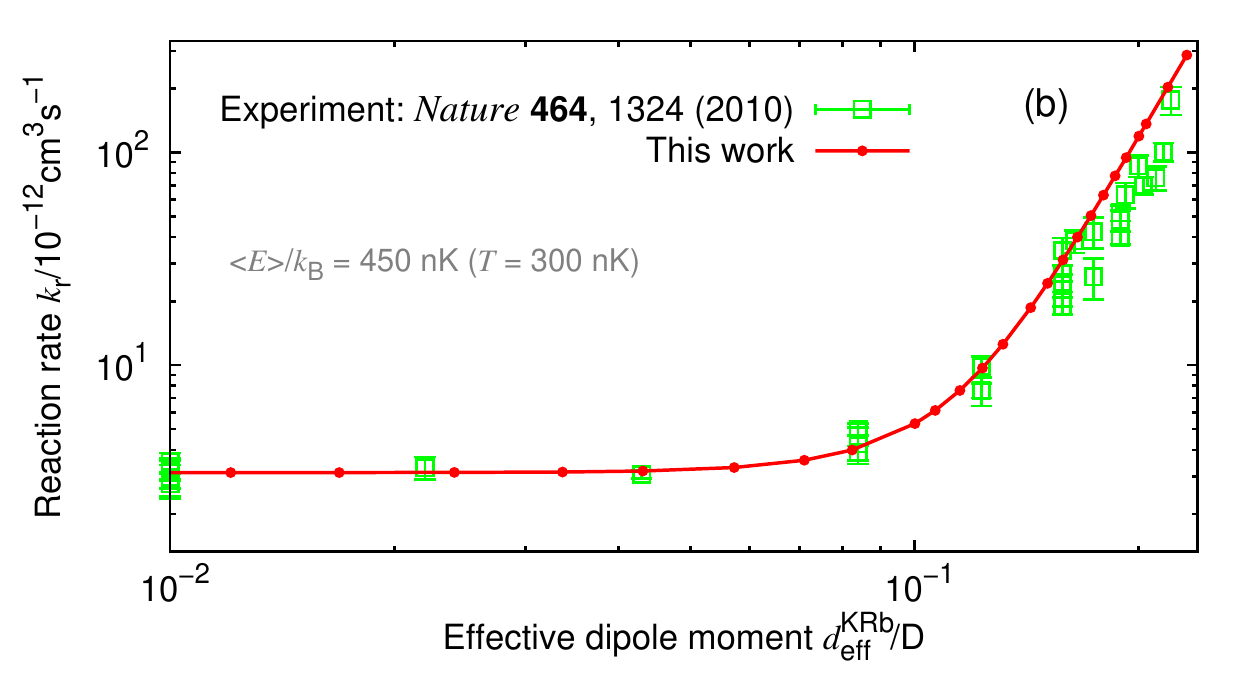}
 \caption{(Color online) Experimental and statistical rates for ultracold
  reactions between ground-state $^{40}$K$^{87}$Rb molecules in: (a) magnetic,
  $F_\mathrm{m}$, and (b) electric, $F_\mathrm{e}$, fields
  ($d^\mathrm{KRb}_\mathrm{eff} \propto F_\mathrm{e}$).}
 \label{fig:kr_KRb+KRb}
\end{figure}

Figure~\ref{fig:kr_KRb+KRb} shows experimental and statistical rates for
ultracold reactions between two ground-state $^{40}$K$^{87}$Rb molecules in
magnetic and electric fields.  Following our previous discussion, the
statistical predictions correspond to those of the WKB and QTM models
\cite{SOspelkaus:10b,K-KNi:10,ZIdziaszek:10a} chosen to evaluate capture
probabilities in the entrance channel(s).  The agreement between experimental
and statistical predictions is hence a consequence of the success of these
models to account for quantum effects in reactions of \emph{undistinguishable
fermions} such as $^{40}$K$^{87}$Rb.  Because of Fermi statistics, such
collisions can only occur with odd partial waves ($L_\alpha = 1, 3,\ldots$) thus
capture in the ultracold channels occurs via tunneling through dynamical
barriers \cite{SOspelkaus:10b,K-KNi:10,MHGdeMiranda:11a}.  These barriers can be
reshaped with applied electric fields, which induce long-range anisotropic
dipolar interactions and lead to different tunneling probabilities for different
$M$ ($M_{L_\alpha}$) values \cite{K-KNi:10,GQuemener:10b,MHGdeMiranda:11a}.
As shown in Fig.~\ref{fig:kr_KRb+KRb}, field effects may induce striking
order-of-magnitude variations in ultracold reaction rates.

It is important to note that quantitative agreement between theoretical and
experimental rates, although a necessary condition, is \emph{not} conclusive
evidence on the statistical nature of a reaction.  Within capture theory,
capture in the entrance channels determines the rates in \emph{any} case with
many more product than reactant channels, which is common in ultracold
experiments.  In any case, Eq.~\eqref{eq:PaX} shows that the key factors that
determine product-state distributions are the \emph{relative} capture
probabilities at different $M$.

\subsection{Statistical product distributions}
\label{sec2:statistical_product_distributions}
Given the striking effect of external fields on ultracold reaction rates, it is
of great interest to assess what effects can fields have on statistical product
distributions.  We focus on translational energy and rotational distributions
for their practical interest, although the formalism can be applied to any
observable of interest.  In particular, translational energy distributions are
likely to play an important role in the initial efforts to measure the products
of ultracold reactions.  At present, reaction experiments have low yields due to
the limited number of reactants, thus translational distributions can be very
valuable because they may provide information on the combined effect of various
DOFs and do not require measuring products in nearly unpopulated states.

\subsubsection{Control of statistical products}
\label{sec3:control_of_statistical_products}
\begin{figure}[!t]
 \includegraphics[width=86mm]{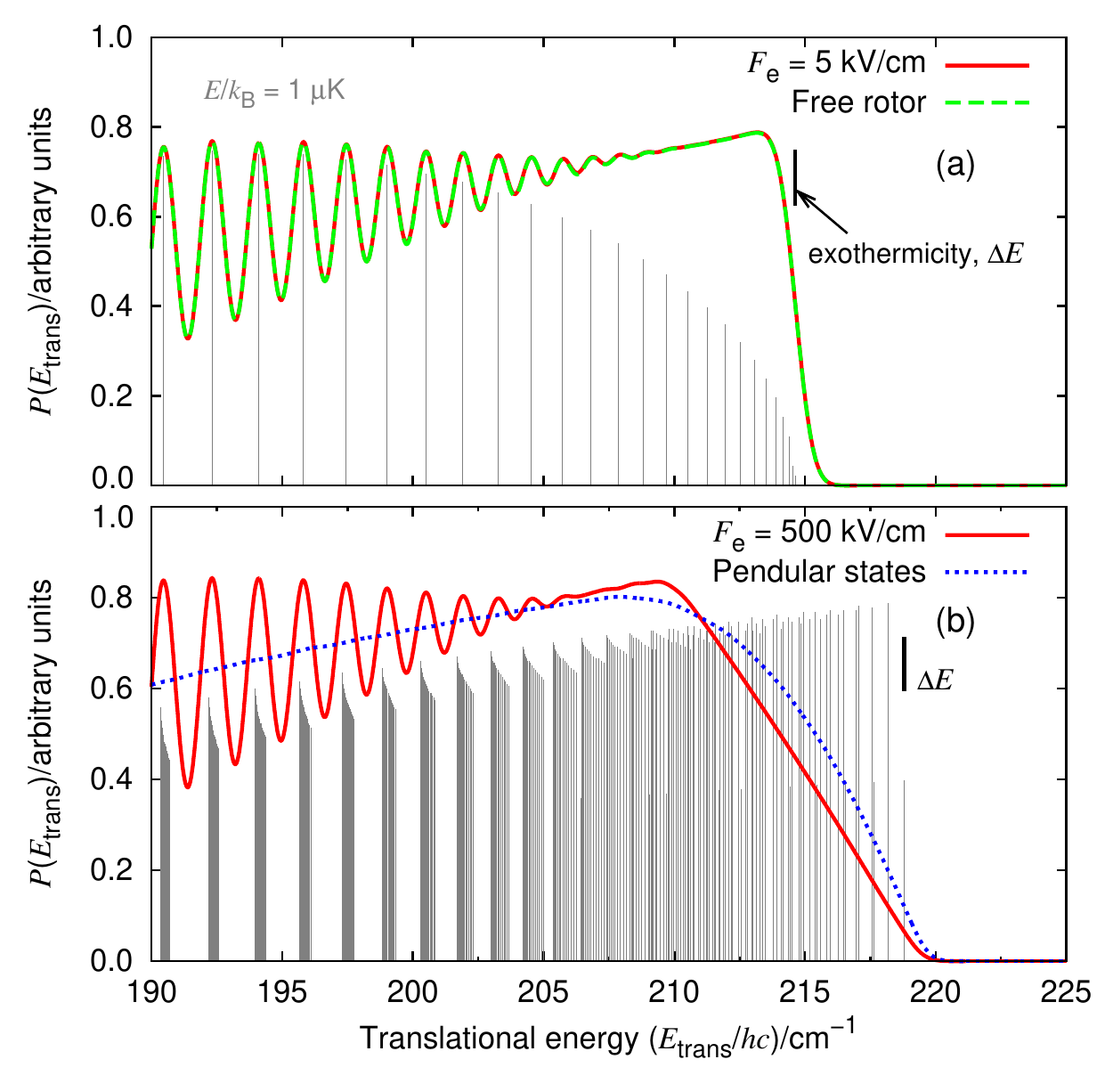}
 \caption{(Color online) Statistical translational energy distributions for the
  ultracold reaction between ground-state $^{40}$K and $^{87}$Rb$_2$ (red,
  solid) at: (a) low, and (b) high electric fields.  The distributions for a
  free rotor (green, dashed) and pendular states (blue, dotted) are shown for
  comparison.}
 \label{fig:PEt_K+Rb2}
\end{figure}
Figure~\ref{fig:PEt_K+Rb2} shows statistical translational energy distributions
for the products of the ultracold reaction $^{40}$K+$^{87}$Rb$_2$ $\rightarrow$
$^{40}$K$^{87}$Rb+$^{87}$Rb at relatively low/high values of an electric field,
with reactants in their ground states.  We focus on the behavior within
30~cm$^{-1}$ of the maximum available kinetic energy and neglect nuclear spins
to keep a manageable number of product states.  The smooth curves are obtained
by convoluting the calculated distributions (sharp peaks) with an ``apparatus''
function (\textit{i.e.}, Gaussian convolution), and are included to emphasize
their overall shape.  The factors determining the form of field-free statistical
distributions have been previously studied \cite{LBonnet:99}.  The results in
Fig.~\ref{fig:PEt_K+Rb2} indicate the degree of control over the
\emph{qualitative} form of product distributions that may be attained if fields
act \emph{directly} on the product energy levels.  There are two main effects:
(1) a change of the spectrum of product states which leads to a change of the
\emph{shape} of the distribution; and (2) a change of the reaction
exothermicity.  Such effects highlight the importance of accounting for fields
when assessing the statistical nature of a reaction, given that statistical
behavior may have different signatures at different field strengths.

Qualitative changes in statistical product distributions are mainly due to the
modification of the product energy levels.  For the 
K+Rb$_2$ reaction, such changes are most noticeable on a reduced energy range
where rotational states are clearly resolved---which is why we restrict
Fig.~\ref{fig:PEt_K+Rb2} to $v = 0$ states, even if KRb($v = 0$--2) products are
energetically available.  A similarly striking qualitative effect on vibrational
distributions would require impractically high electric fields due to the
relatively low electric dipole moment of KRb; this is also why the fields needed
for the strongest effects in the rotational states are very high.  In general,
the key factor determining the energy levels of a polar diatomic rotor is
$d_\mathrm{eff} F_\mathrm{e}/B$ \cite{KVonMeyenn:70}, where $d_\mathrm{eff}$ is
the effective dipole moment, $F_\mathrm{e}$ the applied electric field and $B$
the rotational constant.  In principle, it is thus possible to tune the
rotational distribution of such a product between those of a free rotor and
pendular states simply by tuning the electric field.

\begin{figure}[!t]
 \includegraphics[width=86mm]{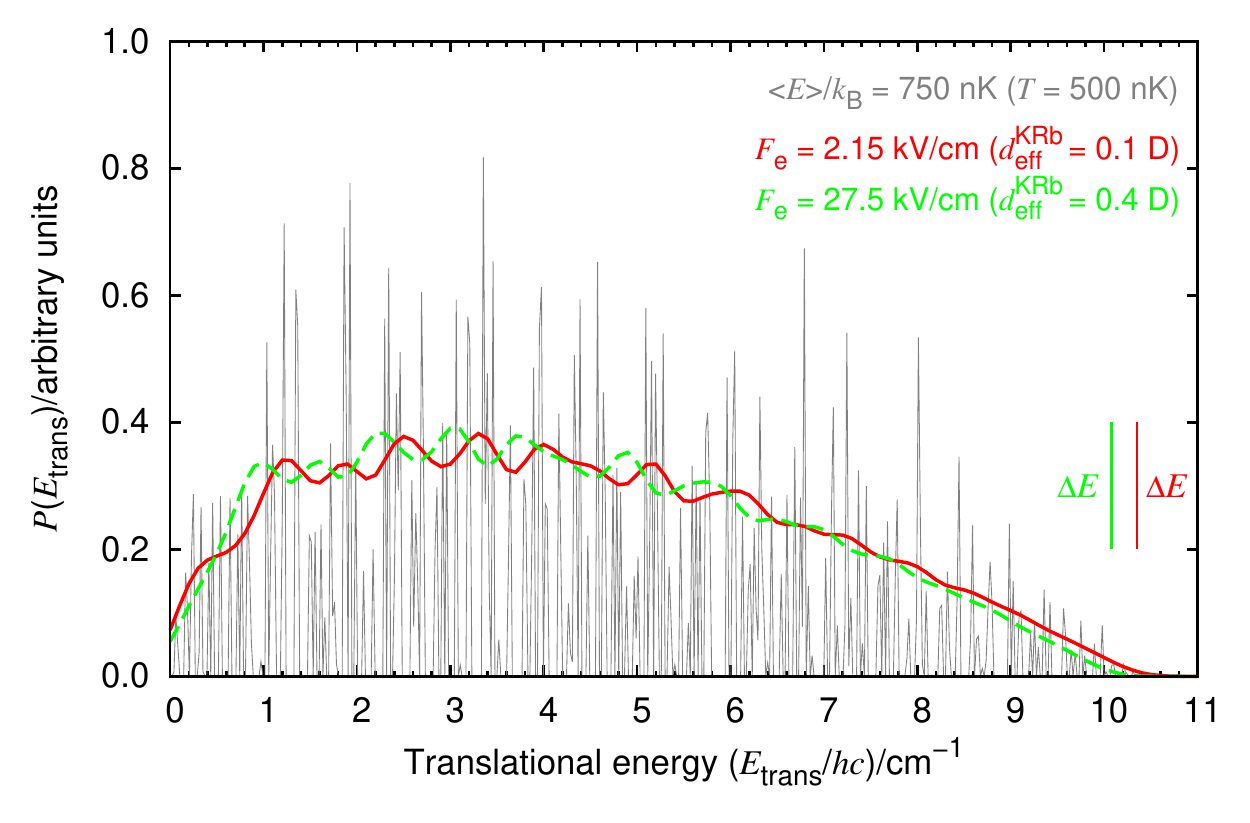}
 \caption{(Color online)  Statistical product translational energy distributions
  for the  ultracold reaction
  $^{40}$K$^{87}$Rb+$^{40}$K$^{87}$Rb $\rightarrow$ K$_2$+Rb$_2$, at different
  electric fields.}
 \label{fig:PEt_KRb+KRb}
\end{figure}
Figure~\ref{fig:PEt_KRb+KRb} shows statistical translational energy
distributions for the prototypical reaction between two ground-state
$^{40}$K$^{87}$Rb molecules.  Calculations include up to hyperfine terms.
Vibrational excitation of the products is prevented because the exothermicity of
the KRb+KRb reaction is low, of about 10 cm$^{-1}$.  In this case, products are
nonpolar species and the field act on the reactant states only.  The main effect
of the field is hence to shift the distributions by changing the reaction
exothermicity, while preserving their qualitative behavior.  In general, the
exothermicity $\Delta E_\alpha \approx \max_\beta{(E_\alpha - E_\beta)}$---given
that $E \approx E_\alpha$---may be increased or decreased with the applied field
depending on whether reactants collide in a low-field-seeking or
high-field-seeking state, which may in principle be used to fully suppress a
reaction \cite{ERMeyer:10}.

Both Figs.~\ref{fig:PEt_K+Rb2} and \ref{fig:PEt_KRb+KRb} show the control that
is feasible on the statistical product distributions of prototypical reactions
using electric fields.  The effect of magnetic fields is different in two ways.
First, large Zeeman splittings in paramagnetic species often require very high
magnetic fields, thus the effects on reaction exothermicities and energy levels
will be less pronounced than for polar species in electric fields.  Secondly,
magnetic fields lift all degeneracies and preserve total parity in the reaction,
which may be used to restrict the product states that become available from
specific initial reactant states.  Effects due to this second feature will be
most pronounced in cases where very few product states are energetically
available.

\subsubsection{Reaction mechanisms}
\label{sec3:reaction_mechanisms}
So far, we assumed that interactions in the intermediate complex mix \emph{all}
DOFs considered, leading to a microcanonical distribution of \emph{all} internal
states.  There exists, however, the possibility that some DOFs are much less
involved and act as ``spectators'', being adiabatically conserved during the
reaction \cite{MQuack:74,RAMarcus:75,GWorry:77}.  Such behavior would have a
distinct signature in the calculated product distributions.  Therefore,
statistical distributions can aid distinguishing between different reaction
mechanisms.  Our formalism can be readily modified to explore these cases by
restricting the sums to product states that fulfill the necessary
constraints---also needed to account for the conservation of quantities other
than $M$.

\begin{figure}[!t]
 \includegraphics[width=86mm]{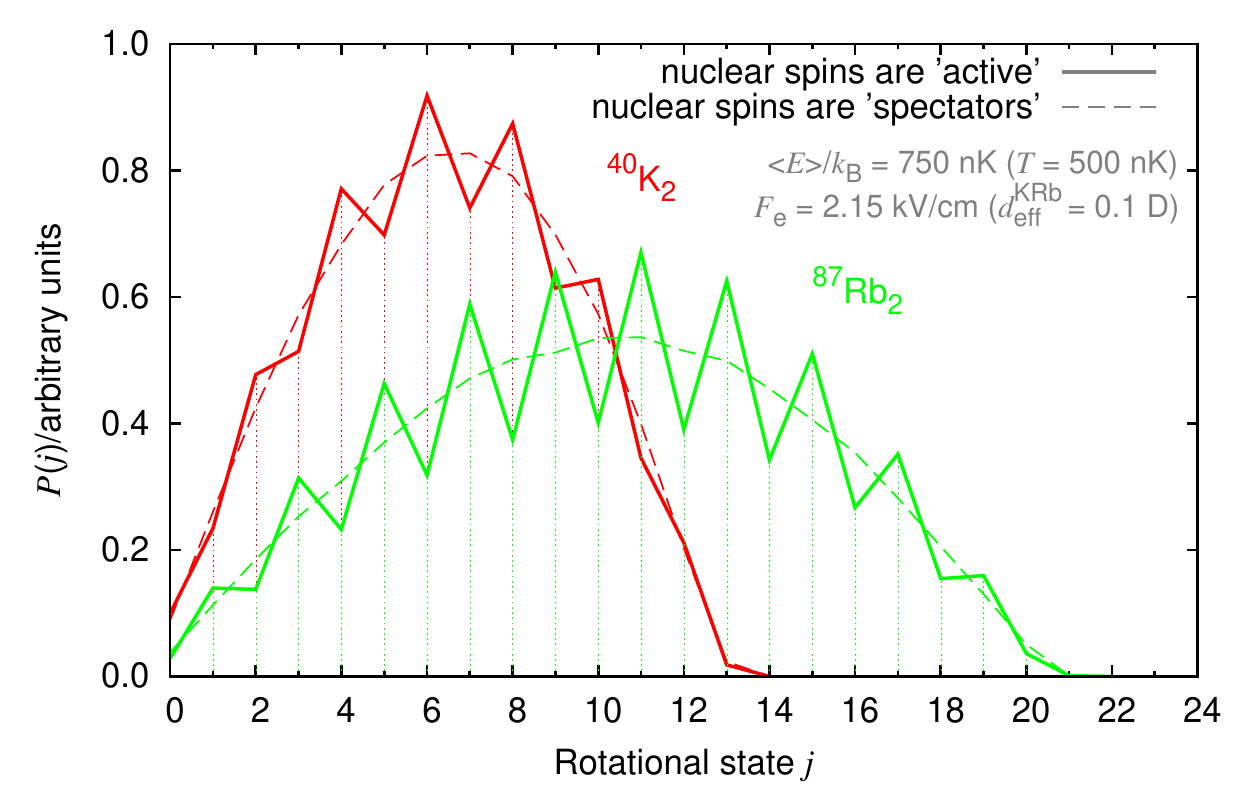}
 \caption{(Color online) Statistical rotational distributions for the products
  $^{40}$K$_2$ (red) and $^{87}$Rb$_2$ (green) of the ultracold reaction between
  two ground-state $^{40}$K$^{87}$Rb molecules in an electric field.
  Solid/dashed lines correspond to the case where nuclear spins are
  conserved/relaxed in the reaction.}
 \label{fig:Pj+-HFS}
\end{figure}

For instance, Mayle \textit{et al.}~\cite{MMayle:12} estimated nuclear
spin-changing probabilities for typical ultracold reactions and predicted that
hyperfine states are likely to change (be preserved) in processes involving
heavier (lighter) species.  Fig.~\ref{fig:Pj+-HFS} shows predicted rotational
distributions for the products of the reaction of two ground-state
$^{40}$K$^{87}$Rb molecules in an electric field, where the \emph{qualitative}
differences between different schemes are clearly demonstrated.  The oscillating
pattern predicted if hyperfine DOFs are ``active'' during the reaction arise
because the products are homonuclear $\Sigma^+_g$-state molecules and exchange
symmetry only allows even/odd values of the total nuclear spin for even/odd
rotational states.  A smooth distribution is predicted if nuclear spins are
spectators.  Both distributions extend to the same maximum rotational number,
fixed by the total available energy.  Statistical tests using predicted curves
as \emph{prior} distributions may be used to assess the involvement of different
DOFs in a reaction, and its variation with varying conditions.

\section{Summary and conclusions}
\label{sec:summary+conclusions}
We have discussed a rigorous statistical formalism for determining
state-to-state cross sections and product distributions in external fields.  We
show that external control of product distributions is possible, although
limited, even for statistical reactions where the ultracold entrance channels
are most ``disconnected'' from the products.  We demonstrate how statistical
predictions can be used to distinguish between different reaction mechanisms.

Product detection in ultracold reaction experiments is still very challenging.
Current traps for neutrals are very shallow which makes in-trap measurements
impractical, while out-of-trap detection is demanding because reactions have low
yields due to the limited number of reactants.  Several leading experimental
groups are trying to achieve the higher densities of reactants that will be
needed and developing new routes for product detection \cite{AHarter:13}.  On
the other hand, ionic products from ion-neutral or Penning-ionization reactions
may be kept for long times using very deep traps and detected with
highly-efficient techniques.

However, a thorough understanding of the physics driving ultracold reactions is
ultimately conditioned by the advances that make possible the measurement of
product-state distributions.  For instance, information on the intermediate
dynamics in ground-state collisions between reactive species (KRb, LiCs,
\textit{etc.}) may be useful in understanding/predicting the dynamics of similar
intermediates in inelastic collisions between nonreactive species (RbCs, NaRb,
\textit{etc.}). In addition, experiments may explore reactions in mixed samples
of homonuclear dimers such as Rb$_2$+Cs$_2$ $\rightarrow$ 2RbCs in an attempt to
better understand the inverse processes. All these considerations add to the
relevance of alkali-metal dimers that are considered as less topical mostly
because of their ground-state reactivity at ultracold temperatures
\cite{PSZuchowski:10b,MTomza:13a}.

\begin{acknowledgments}
The authors are grateful to Goulven Qu\'em\'ener and Piotr S.\ \.{Z}uchowski for
discussions and comments on the manuscript.  MLGM acknowledges funding from the
7th European research program FP7/2007-2013 under grant agreement No.~330623.
\end{acknowledgments}

\appendix*
\section{Calculation details}
\label{app:calculation_details}
This Appendix describes the calculation of quantum states and capture
probabilities required for applying Eqs.~\eqref{eq:sigma_ab,r} and
\eqref{eq:PaX} to the study of $^{40}$K+$^{87}$Rb$_2$ $\rightarrow$ KRb+Rb and
$^{40}$K$^{87}$Rb+$^{40}$K$^{87}$Rb $\rightarrow$ K$_2$+Rb$_2$ reactions.

\subsubsection{Quantum states}
\label{sec3:quantum_states}
The Hamiltonian for the isolated alkali atoms, $^{40}$K($^2$S) and
$^{87}$Rb($^2$S), is in general split into hyperfine and Zeeman contributions
$\hat{\mathcal{H}} = \hat{\mathcal{H}}_\mathrm{hf} +
\hat{\mathcal{H}}_\mathrm{Z}$.  $\hat{\mathcal{H}}_\mathrm{hf} =
a\,\hat{\imath}\cdot\hat{s}$, with $a$ the atomic hyperfine coupling constant,
$\hat{\imath}$ and $\hat{s}$ the nuclear and electronic spin operators.
$\hat{\mathcal{H}}_\mathrm{Z} = g_\mathrm{S} \mu_\mathrm{B} \hat{s} \cdot
\hat{F}_\mathrm{m} - g_i \mu_\mathrm{N} \hat{\imath} \cdot \hat{F}_\mathrm{m}$,
where $g_\mathrm{S} \approx 2$ and $g_i$ are the electron and nuclear
$g$-factors, $\mu_\mathrm{B}$ and $\mu_\mathrm{N}$ the Bohr and nuclear
magnetons, and $\hat{F}_\mathrm{m}$ the external magnetic field operator.  The
matrix elements for this Hamiltonian are taken from the literature,
\textit{cf.}~Ref.~\cite{MLGonzalez-Martinez:13c}.  Quantum states are calculated
from the diagonalization of the resulting Hamiltonian matrix for a specific
value of the external fields, using the parameters from NIST atomic database
\cite{NIST:atomic}.

The quantum states for $^{40}$K$_2$($^1\Sigma^+_g$),
$^{87}$Rb$_2$($^1\Sigma^+_g$) and $^{40}$K$^{87}$Rb($^1\Sigma^+$), in their
vibrational ground states, are computed using the formalism in
Refs.~\cite{JAldegunde:08,JAldegunde:09}.  In general, the molecular Hamiltonian
is split into rotational, hyperfine, Zeeman and Stark terms as
$\hat{\mathcal{H}} = \hat{\mathcal{H}}_\mathrm{rot} +
\hat{\mathcal{H}}_\mathrm{hf} + \hat{\mathcal{H}}_\mathrm{Z} +
\hat{\mathcal{H}}_\mathrm{S}$, which read
\begin{eqnarray}
 \hat{\mathcal{H}}_\mathrm{rot} &=& B\hat{n}^2 - D\hat{n}^2\hat{n}^2,\nonumber\\
 \hat{\mathcal{H}}_\mathrm{hf} &=& \sum^2_{k = 1} \hat{V}_k:\hat{Q}_k
                       + \sum^2_{k = 1} c_k \hat{n} \cdot \hat{\imath}_k
                      + c_3 \hat{\imath}_1 \cdot \mathrm{T} \cdot \hat{\imath}_2
                           + c_4 \hat{\imath}_1 \cdot \hat{\imath}_2,\nonumber\\
 \hat{\mathcal{H}}_\mathrm{Z} &=&
     -g_\mathrm{r} \mu_\mathrm{N} \hat{n} \cdot \hat{F}_\mathrm{m}
     -\sum^2_{k=1} g_k \mu_\mathrm{N}
                \hat{\imath}_k \cdot \hat{F}_\mathrm{m} (1-\sigma_k),\nonumber\\
 \hat{\mathcal{H}}_\mathrm{S}   &=& -\hat{d} \cdot \hat{F}_\mathrm{e}.
\end{eqnarray}
Here, $\hat{n}$ is the operator of the rotational angular momentum, while
$\hat{\imath}_k$ ($k = 1, 2$) are those of the nuclear spins of the individual
atoms.  $B$ and $D$ are the rotational and centrifugal distortion constants (we
neglect the latter, following Refs.~\cite{JAldegunde:08,JAldegunde:09}).  The
first term in the hyperfine Hamiltonian corresponds to the interaction of the
electric quadrupole moment of each nucleus, $Q_k$, with the electric field
gradient it experiences, $V_k$.  The other three hyperfine terms correspond to
the nuclear-spin rotation for each nucleus (with associated coupling constants
$c_1$ and $c_2$), the tensor ($c_3$) and scalar ($c_4$) interactions between the
nuclear spins.  $\mathrm{T}$ is a tensor describing the angular dependence of
the direct spin-spin interaction and the anisotropic part of the indirect
spin-spin interaction \cite{DLBryce:03}.  The Zeeman Hamiltonian involves a
rotation term with rotation $g$-factor $g_\mathrm{r}$, and nuclear terms
including the isotropic part of the nuclear shielding tensor, $\sigma$.  The
Stark Hamiltonian is only relevant for KRb, and involves the electric dipole
moment $\hat{d}$ and electric field $\hat{F}_\mathrm{e}$ operators.  All matrix
elements are given in the appendices of
Refs.~\cite{JAldegunde:08,JAldegunde:09}; the molecular parameters for
$^{40}$K$_2$ and $^{87}$Rb$_2$ are reported in Table~I of
Ref.~[\onlinecite{JAldegunde:09}], while those for $^{40}$K$^{87}$Rb are in
Table~V of Ref.~[\onlinecite{JAldegunde:08}].

Only $^{40}$K$_2$($^1\Sigma^+_g,\,v = 0$) and $^{87}$Rb$_2$($^1\Sigma^+_g,\,v =
0$) products are populated in the KRb+KRb reaction, with a zero-field
exothermicity of $-$10.355~cm$^{-1}$ \cite{SOspelkaus:10b}.  The zero-field
exothermicity of the $^{40}$K+$^{87}$Rb$_2$ reaction is $-$214.617~cm$^{-1}$
\cite{SOspelkaus:10b}, and does not exceed $-$220~cm$^{-1}$ for the electric
fields considered.  Hence, only vibrational states $v = 0$--2 of the KRb
products are energetically accessible \cite{CAmiot:00}.  For the latter
reaction, we neglect the couplings between states corresponding to different
vibrational manifolds and the variation of molecular constants with the
vibrational state $v$.  Different vibrational states are added a vibrational
energy of $E_\mathrm{vib} = \hbar w (v + \frac{1}{2})$, with
$w = 75.85$~cm$^{-1}$ for KRb \cite{CAmiot:00}.

\subsubsection{Capture probabilities}
\label{sec3:capture_probabilities}
Three different capture models were used in the calculations, two quantum models
for the capture of ultracold reactants: (1) a WKB approximation for collisions
in magnetic fields, and (2) the adiabatic variant of QTM by Qu\'em\'ener and
Bohn \cite{GQuemener:10b} for collisions in electric fields; and one for the
capture of products: (3) a semiclassical model based on PST
\cite{JCLight:64,PPechukas:65,EENikitin:65,CKlotz:71}.

In our calculations, we approximate the capture probabilities for the ultracold
reactants by the tunneling probability through dynamical barriers in the
entrance channels.  All dispersion coefficients are taken form \emph{ab initio}
data \cite{PSZuchowski:13a,MKosicki:pc14}

\paragraph*{WKB capture model.}
The capture probability is calculated using the WKB expression for the
transmission probability through a centrifugal barrier
\begin{widetext}
 \begin{equation}
  p^M_\gamma(E, F_\mathrm{m}) =
              \exp{\left\{-\frac{2}{\hbar}\int^{R_\mathrm{max}}_{R_\mathrm{min}}
    \sqrt{ 2\mu_\gamma \left[\frac{L_\gamma(L_\gamma+1)\hbar^2}{2\mu_\gamma R^2}
               - \frac{C_{6,\gamma}}{R^6} - (E-E_\gamma)\right]}\, dR \right\}},
 \end{equation}
\end{widetext}
where $R_\mathrm{min}$ and $R_\mathrm{max}$ are the classical turning points
(zeroes of the factor within square brackets), which are determined numerically.
$\mu_\gamma$ and $C_{6,\gamma}$ are the relevant reduced mass and long-range
dispersion coefficient, while $L_\gamma$ is the orbital angular quantum number.
The dependence of this probability with the applied magnetic field,
$F_\mathrm{m}$, arises from that of $E_\gamma$ (and thus of $R_\mathrm{min}$ and
$R_\mathrm{max}$).

\paragraph*{QTM capture model.}
The model was developed by Qu\'em\'ener and Bohn \cite{GQuemener:10b} to account
for the strong effect of the electric dipole-dipole interaction on the dynamical
barriers.  The capture probability is calculated from their model for the
transition probability, Eq.~(14) in Ref.~\cite{GQuemener:10b},
\begin{equation}
  p^M_\gamma(E, F_\mathrm{e}) = q \times
          \left(\frac{E-E_\gamma}{V_{\mathrm{b}, \gamma}}\right)^{L_\gamma+1/2},
\end{equation}
where $q \approx 2/3$ is a correction factor and $V_{\mathrm{b}, \gamma}$ the
height of the dynamical barrier \cite{GQuemener:10b,K-KNi:10}.  In this case,
the dependence on the electric field $F_\mathrm{e}$ arises from that of
$V_{\mathrm{b}, \gamma}$.  The barrier height is obtained from the
diagonalization of the long-range interaction including the electric
dipole-dipole term.  In the orbital angular momentum basis, the matrix elements
read
\begin{eqnarray}
\label{eq:<LML|V|L'ML>}
 &&\hspace{-62mm}
 \langle L_\gamma M_{L_\gamma} | V(R) | L'_\gamma M_{L_\gamma} \rangle
   \nonumber\\
   = \delta_{L_\gamma, L'_\gamma}
 \left[\frac{\hbar^2 L_\gamma(L_\gamma+1)}{2\mu_\gamma R^2} -
                                \frac{C_{6,\gamma}}{R^6} \right]
    &&- \frac{C_{3,\gamma}}{R^3},
\end{eqnarray}
with
\begin{eqnarray}
 C_{3,\gamma} &=& 2 d^\gamma_{\mathrm{eff}, 1} d^\gamma_{\mathrm{eff}, 2}
                  (-1)^{M_{L_\gamma}}
         \left[(2L_\gamma+1)(2L'_\gamma+1)\right]^{1/2} \nonumber\\
 && \times
    \left(\begin{array}{ccc} L_\gamma & 2 & L'_\gamma \\
                             0 & 0 & 0  \end{array}\right)
    \left(\begin{array}{ccc} L_\gamma    & 2 & L'_\gamma  \\
                            -M_{L_\gamma} & 0 & M_{L_\gamma} \end{array}\right),
\end{eqnarray}
where $d^\gamma_{\mathrm{eff}, k}$ ($k=1,2$) are the effective/induced electric
dipole moments for each species and $\left(:::\right)$ represents a Wigner 3-$j$
symbol.  The barrier height $V_{\mathrm{b}, \gamma}$ is the smallest eigenvalue
of the matrix defined by Eq.~\eqref{eq:<LML|V|L'ML>}, which is diagonalized for
each $M_{L_\gamma}$ ($M$) using $L_\gamma = 1, 3, \ldots, 9$.

\paragraph*{PST capture model.} This model is used for capture in the product
channels only, where the kinetic energy is relatively large and the effect of
a few near-threshold states below dynamical barriers may be neglected.  The
capture probabilities are computed as
\begin{equation}
 p^M_\gamma(E, F) =
 \left\{\begin{array}{l} 1,\; \mathrm{if}\,E-E_\gamma\ge V_{\mathrm{b},\gamma}\\
                         0,\; \mathrm{otherwise,} \end{array}\right.
\end{equation}
where the height of the centrifugal barrier is
\begin{equation}
 V_{\mathrm{b},\gamma} = \sqrt{\frac{\left[L_\gamma(L_\gamma+1)\hbar^2\right]^3}
                              {54 \mu^3_\gamma C_{6, \gamma}}}.
\end{equation}
$p^M_\gamma$ dependence on the field(s) arises from that of $E_\gamma$.


\end{document}